\documentclass[aps,prd,floatfix,nofootinbib,superscriptaddress,preprint]{revtex4-1}
\usepackage{epsfig}
\usepackage{amstext,amssymb}
\usepackage{amsmath}
\usepackage{graphicx}
\usepackage{dcolumn}
\usepackage{color}
\usepackage{slashed} 
\usepackage{float}
\usepackage[caption = false]{subfig}

\def\({\left(}
\def\){\right)}

\begin{document}

\title{\boldmath $B-L$ violating nucleon decays as a probe of leptoquarks and implications for baryogenesis}
\author{Chandan Hati}
\email{c.hati@tum.de} 
\affiliation{Physik Department T70, Technische Universit\"at M\"unchen,\\
James-Franck-Stra{\ss}e 1, D-85748 Garching, Germany}
\affiliation{Laboratoire de Physique de Clermont (UMR 6533) --- CNRS/IN2P3,\\ 
Univ. Clermont Auvergne, 4 Av. Blaise Pascal, F-63178 Aubi\`ere Cedex,
France }
\author{Utpal Sarkar}
\email{utpal@phy.iitkgp.ernet.in}
\affiliation{Department of Physics, Indian Institute of Technology Kharagpur, Kharagpur 721302, India}

\begin{abstract}
We study the effective $B-L$ violating couplings for scalar and vector leptoquarks which can naturally induce dimension seven $B-L$ violating operators leading to very interesting nucleon decay modes such as $n \rightarrow e^- \pi^+, e^-K^+,\mu^- \pi^+, \mu^-K^+$ and $p \rightarrow \nu \pi^+$. This opens a new window to probe the nature and couplings of the scalar and vector leptoquarks in an ultraviolet model independent way which can provide an orthogonal probe for scalar and vector leptoquark solutions to the recent anomalous $B$-decay data. Furthermore, we also discuss how these $B-L$ violating interactions can also pave a new way to understand the origin of matter-antimatter asymmetry of the Universe. 
\end{abstract}

\maketitle

\section{Introduction}
The renormalisable Standard Model (SM) Lagrangian interactions cannot violate baryon number, and consequently, baryon number violation can only arise through effective higher dimensional operators with dimension six or higher \cite{Weinberg:1979sa,Wilczek:1979hc,Abbott:1980zj}, making baryon number violation a very sensitive probe of new physics beyond the SM. The leading dimension six operators can naturally be generated in many grand unified theory (GUT) embedding of the SM and they conserve baryon number minus lepton number $(B-L)$ symmetry leading to proton decay modes such as $p \rightarrow e^+ \pi^0$, $p \rightarrow \mu^+ \pi^0$ and $p \rightarrow \overline{\nu} K^+$, which have been the primary focus of the recent experimental searches.  On the other hand, $d=7$ operators obeying the selection rule $\Delta (B-L) = -2$  \cite{Weinberg:1980bf,Weldon:1980gi} leading to interesting decay modes such as $n \rightarrow e^- \pi^+, e^-K^+,\mu^- \pi^+, \mu^-K^+$ and $p \rightarrow \nu \pi^+$ have received considerably less attention. These $B-L$ violating decay modes have been discussed in the context of $G_{\text{Pati-Salam}}$, and $SO(10)$ GUT theories in Ref. \cite{Pati:1983zp} and in the context of $SO(10)$ GUT theories in Refs. \cite{Babu:2012vb,Babu:2012iv}. In these studies, driven by the GUT theory motivations, the mass scales of the mediating particles have been taken to be very heavy $\sim 10^{11}-10^{13}$ GeV. In this work we point out that these $B-L$ violating nucleon decay modes provide a novel way to probe the nature and couplings of the scalar and vector leptoquarks in a model independent way \footnote{For discussion of $\Delta (B)=2 $ transitions involving trilinear di-quark interactions (forbidding leptoquark-like couplings) that can be probed in neutron-antineutron transitions and di-nucleon decays see Ref. \cite{Baldes:2011mh}.}. This not only makes these $B-L$ violating nucleon decay modes extremely interesting for current and future experimental searches, but also gives way to a new type of mechanism to understand matter-antimatter asymmetry of the Universe where simultaneous baryon and lepton number violation gives rise to a $B-L$ violating asymmetry, which unlike $B-L$ conserving asymmetry, does not get washed out by $B+L$ violating sphaleron interactions near the electroweak scale \cite{Kuzmin:1985mm}. Furthermore, this also motivates the experimental search for $B-L$ violating nucleon decay modes as an orthogonal probe for TeV scale scalar and vector leptoquark solutions to the $B-$decay anomalies, which are persistent with new data from the $B-$factories \cite{Lees:2012xj,Lees:2013uzd, Huschle:2015rga,Adachi:2009qg, Bozek:2010xy, Aaij:2015yra,Hirose:2016wfn, Abdesselam:2019dgh, Aaij:2019wad, Aaij:2017vbb, Abdesselam:2019wac, Aaij:2015esa, Wehle:2016yoi}, and are currently one of the strongest hints of new physics beyond the standard model. To this end, the recent measurements of the ratio of branching fractions of $B\to K(K^\ast)\ell\ell$ decays into di-muons over di-electron modes $R_{K^{(\ast)}}$ and the ratio of branching fractions of $B \to  D^{(*)} \ell^- \bar\nu$ decays into tau over other lepton modes $R_{D^{(\ast)}}$ are of particular interest because in these ratios the hadronic uncertainties cancel and consequently, these observables are sensitive to lepton flavour universality (LFU) violating new physics.

The latest measurement of $R_K$ for the dilepton invariant mass squared bin
$[1.1,6]~\text{GeV}^{2}$~\cite{Aaij:2019wad} shows a $2.5\,\sigma$
deviation falling short of the SM
prediction~\cite{Bordone:2016gaq,Capdevila:2017bsm}. Similarly, the measurement
of $R_{K^*}$~\cite{Aaij:2017vbb} display $2.3\,\sigma$ and
$2.6\,\sigma$ deviations, below the SM predictions for the dilepton
invariant mass squared bins $[0.045, 1.1]~\text{GeV}^{2}$ and
$[1.1, 6]~\text{GeV}^{2}$, respectively~\cite{Bordone:2016gaq,Capdevila:2017bsm}. Further, neutral current anomalies have also emerged in the observable $\Phi \equiv d \text{BR}(B_s\to\phi\mu\mu)/ dm_{\mu\mu}^2$, in a similar kinematic regime
($m_{\mu\mu}^2\in[1,6]\,{\rm GeV}^2$)~\cite{Aaij:2015esa,Altmannshofer:2014rta,Straub:2015ica},
amounting to a deviation of about $3\,\sigma$. Deviations from
the SM expectations have also been reported in the angular observable
$P_5^{\prime}$ for the $B \to K^\ast \ell^+ \ell^-$ decay. On the other hand, the current measured values of $R_D$~\cite{Amhis:2016xyh, Abdesselam:2019dgh} and $R_{D^\ast}$~\cite{Aaij:2015yra,Hirose:2016wfn,Amhis:2016xyh,Abdesselam:2019dgh} exceed the expected SM values by about $1.4\,\sigma$ and $2.5\,\sigma$ respectively~\cite{Bigi:2016mdz,Bigi:2017jbd}, and when combined they lead to a deviation of $3.1\,\sigma$ from the SM
prediction~\cite{Ligeti:2016npd,Crivellin:2016ejn,
  Amhis:2016xyh}.

 These anomalies have resulted in a plethora of very interesting studies, among which two very popular scenarios which can address the above anomalies are the extensions of the SM with scalar leptoquarks and vector leptoquarks \cite{Dorsner:2016wpm,Alguero:2019ptt,Aebischer:2019mlg,Ciuchini:2019usw,Datta:2019zca,Arbey:2019duh,Shi:2019gxi,Bardhan:2019ljo,Alok:2019ufo,Alok:2017qsi,Ghosh:2014awa,Glashow:2014iga,Bhattacharya:2014wla,Freytsis:2015qca,Ciuchini:2017mik,Hiller:2014yaa,Gripaios:2014tna,Sahoo:2015wya,Varzielas:2015iva,Alonso:2015sja,Bauer:2015knc,Hati:2015awg,Fajfer:2015ycq,Das:2016vkr,Becirevic:2016yqi,Sahoo:2016pet,Cox:2016epl,Crivellin:2017zlb,Becirevic:2017jtw,Cai:2017wry,Dorsner:2017ufx,Greljo:2018tuh,Sahoo:2018ffv,Becirevic:2018afm,Hati:2018fzc,deMedeirosVarzielas:2018bcy,Aebischer:2018acj,deMedeirosVarzielas:2019okf,Yan:2019hpm,Bigaran:2019bqv,Popov:2019tyc,Deshpand:2016cpw,Altmannshofer:2017poe,Das:2017kfo,Earl:2018snx,Trifinopoulos:2018rna,Trifinopoulos:2019lyo,Assad:2017iib,Buttazzo:2017ixm,Calibbi:2017qbu,Bordone:2017bld,Blanke:2018sro,Bordone:2018nbg,Kumar:2018kmr,Angelescu:2018tyl,Balaji:2018zna,Fornal:2018dqn,Baker:2019sli,Cornella:2019hct,DaRold:2019fiw,Barbieri:2016las,Hati:2019ufv}. Vector leptoquark couplings have the potential to simultaneously explain both $R_{K^{(\ast)}}$ and $R_{D^{(\ast)}}$ \cite{Buttazzo:2017ixm}, however a minimal $G_{\text{Pati-Salam}}$ GUT embedding of such vector leptoquark is inconsistent with very strong bounds from various flavour violating processes. Consequently, non-minimal extensions of such embedding is required to avoid such bounds \cite{Assad:2017iib,Buttazzo:2017ixm,Calibbi:2017qbu,Bordone:2017bld,Blanke:2018sro,Bordone:2018nbg,Kumar:2018kmr,Angelescu:2018tyl,Balaji:2018zna,Fornal:2018dqn,Baker:2019sli,Cornella:2019hct,DaRold:2019fiw,Barbieri:2016las,Hati:2019ufv}. On the other hand, single scalar leptoquark cannot simultaneously address $R_{K^{(\ast)}}$ and $R_{D^{(\ast)}}$ data without violating constraints from various flavour violating processes and consequently attempts have been made to use more than one leptoquark and to embed them in minimal GUT frameworks to such end \cite{Dorsner:2017ufx,Dorsner:2017wwn}. In this work we will primarily focus on the scalar leptoquark states tabulated in Table. I and the vector leptoquarks $U_1: (\bf{3},\bf{1},2/3)$ and $\tilde{V_2}: (\overline{\bf{3}},\bf{2},-1/6)$.
\begin{table}[h]
\centering
\begin{tabular}{|c|c|c|c|c|}
\hline
LQ & $G_{\text{SM}}$ & $SU(5)$ & $G_{\text{Pati-Salam}}$ & $SO(10)$\\
\hline 
$S_1$ & $(\overline{\bf{3}},\bf{1},1/3)$ & $\overline{\bf{5}},\, \overline{\bf{45}},\, \overline{\bf{50}}$ & (\bf{1},\bf{1},\bf{6}),\, (\bf{1},\bf{3},$\bf{\overline{10}}$) & $\bf{10},\,\bf{120},\,\overline{\bf{126}}$\\
$S'_1$ & $(\overline{\bf{3}},\bf{1},-2/3)$ & $ \bf{10}$ & $(\bf{1},\bf{3},\bf{6})$ & $\bf{120}\,\overline{\bf{126}}$\\
$\tilde{S}_1$ & $(\overline{\bf{3}},\bf{1},4/3)$ & $ \bf{45}$ &  $(\bf{1},\bf{3},\bf{6})$ & $\bf{120}\,\overline{\bf{126}}$\\
$\tilde{R}_2$ & $(\bf{3},\bf{2},1/6)$ & $\bf{10},\,\bf{15}$ & $(\bf{2},\bf{2},\bf{15})$ & $\bf{120},\,\overline{\bf{126}}$\\
$R_2$ & $(\bf{3},\bf{2},7/6)$ & $\overline{\bf{45}},\, \overline{\bf{50}}$ & $(\bf{2},\bf{2},\bf{15})$ & $\bf{120},\,\overline{\bf{126}}$\\
$S_3$ & $(\overline{\bf{3}},\bf{3},1/3)$ & $\overline{\bf{45}}$ & $(\bf{3},\bf{1},\bf{6})$  & $\bf{120},\,\overline{\bf{126}}$ \\
\hline
\end{tabular}
\caption{\label{tab:1} Transformation properties of scalar leptoquarks under the SM gauge group $G_{\text{SM}}\equiv SU(3)_c \times SU(2)_L \times U(1)_Y $ and the relevant $SU(5)$, $G_{\text{Pati-Salam}}\equiv SU(2)_L \times SU(2)_R \times SU(4)_c$, $SO(10)$ representations.}
\end{table} 

The plan for rest of this paper is as follows. In section \ref{sec2}, we discuss the effective leptoquark couplings and $B-L$ violating operators with $d=7$ and provide the relevant diagrams and estimation for the lifetime for nucleon decay modes. In section \ref{sec3}, we discuss possible UV completions for these interactions and the possible origins of the $B-L$ violating couplings in the context of these UV completions. In section \ref{sec4}, we discuss the possible connection between the $B$-decay anomalies and the $B-L$ violating nucleon decay rates. In section \ref{sec5}, we explore the implications of these $B-L$ violating interactions for baryogenesis and the possible correlation between the final baryon asymmetry and the nucleon decay lifetime. Finally, in section \ref{sec5} we summarise and make our concluding remarks.
\section{Effective leptoquark couplings and $B-L$ violating operators with $d=7$}{\label{sec2}}
The dimension seven $B$--violating effective operators invariant under the SM gauge group which follow the selection rule $\Delta(B-L) = +2$ and mediate $(B-L)$ violating nucleon decays are given by \cite{Weinberg:1980bf,Weldon:1980gi}
\begin{eqnarray}
{\cal {O}}_1 &=& (Q_i Q_j)(d^c L_k)^*H^*_l \epsilon_{ij} \epsilon_{kl},
{\cal {O}}_2 = (Q_i Q_j) (d^c L_j)^*H^*_i , \nonumber \\
{\cal {O}}_3 &=& (d^c d^c)^* (Q_i e^c) H^*_i,~~~~~~~~
{\cal {O}}_4 = (d^c d^c)^* (u^c L_i)^* H^*_j \epsilon_{ij} \nonumber \\
{\cal {O}}_5 &=& (d^c u^c)^* (d^c L_i)^* H^*_j \epsilon_{ij},~~~
{\cal {O}}_6 = (d^c d^c)^*(d^c L_i)^* H_i, \nonumber\\
{\cal {O}}_7 &=& (d^c D_\mu d^c)^*(\overline{L}_i \gamma^\mu Q_i),~~~~
{\cal {O}}_8 = (d^c D_\mu L_i)^*(\overline{d^c} \gamma^\mu Q_i), \nonumber \\
{\cal {O}}_9 &=& (d^c D_\mu d^c)^* (\overline{d^c} \gamma^\mu e^c)~,
\label{dim7}
\end{eqnarray}
where $i,j,k,l$ are $SU(2)_L$ indices, and all the fermion fields are written in terms of left handed spinors. We will first show that some of these operators can arise naturally in the presence of effective trilinear couplings involving two scalar (vector) leptoquarks and the SM Higgs doublet. Interestingly, at the effective level the coupling coefficient of these trilinear interactions have mass dimension unity and is proportional to the $B-L$ breaking scale, which can in principle be decoupled from the gauge coupling unification scale in a given UV complete model. We will first treat the problem at an effective level where we will only fix the effective mass scales corresponding to the coupling coefficient of this trilinear interactions and the mass scales of the leptoquark states to show that these operators can give rise to nucleon decay rates which can be probed at the current and future experimental nucleon decay search facilities. The relevant effective trilinear scalar interactions invariant under the SM gauge group $G_{\text{SM}}\equiv SU(3)_c \times SU(2)_L \times U(1)_Y $ are given by
\begin{eqnarray}
\label{eq:trilinear}
\mathcal{L}_\mathrm{scalar} \supset
 - \lambda_1 \tilde{R}_{2}^{\dagger} H S^{\dagger}_1 - \tilde{\lambda}_2 \tilde{R}^{\dagger}_{2} H^{\dagger} {S'}_1^{\dagger} - \lambda_2 R_{2}^{{\dagger}} H {S'}_1^{\dagger}  
- \lambda_3 \tilde{R}_{2}^{\dagger\,i} (\vec{\tau}. S_{3})^{\dagger\;ij} H^{k} +\textrm{h.c.}\; ,
\end{eqnarray}
where $i,j,k$ are $SU(2)_L$ indices. To see the $B-L$ assignments of the scalar leptoquark states and the above couplings let us write down the Yukawa interactions for the relevant scalar leptoquarks at an effective level--- 
\begin{eqnarray}
\label{eq:yukawa1}
\mathcal{L}_\mathrm{Y_{S_1}} \supset&& -(y^{1}_{S_1} Q_{L}^{i}L_{L}^{j}\epsilon^{ij} + y^{2}_{S_1}\epsilon^c u^{c}_L d^{c}_L) S_{1} 
 -(y^{3}_{S_1} \epsilon^c Q_{L}^{i}Q_{L}^{j}\epsilon^{ij} + y^{4}_{S_1} u^{c}_L e^{c}_L + y^{5}_{S_1} d^{c}_L \nu^{c}_L) S_{1}^{\dagger} \\
\label{eq:yukawa2}
\mathcal{L}_\mathrm{Y_{S'_1,\tilde{S}_1}} \supset&& -y^{1}_{S'_1} u_{L}^{c}\nu_{L}^{c} {S'}_1^{\dagger} - y^{2}_{S'_1}\epsilon^c d^{c}_L d^{c}_L {S'}_{1}
 -y^{1}_{\tilde{S}_1} d_{L}^{c}e_{L}^{c} \tilde{S}_1- y^{2}_{\tilde{S}_1}\epsilon^c u^{c}_L u^{c}_L \tilde{S}_{1}^{\dagger}\\
\label{eq:yukawa3}
\mathcal{L}_\mathrm{Y_{\tilde{R}_2,R_2}} \supset&& -y^{1}_{\tilde{R}_2}d^{c}_L L_{L}^{i}\tilde{R}_{2}^{j}\epsilon^{ij}-y^{2}_{\tilde{R}_2}\nu^{c}_L Q_{L}^{i}\tilde{R}_{2}^{\dagger\;j}\epsilon^{ij}
 -y^{1}_{R_2}u^{c}_L L_{L}^{i}R_{2}^{j}\epsilon^{ij}-y^{2}_{R_2} e^{c}_L Q_{L}^{i}R_{2}^{\dagger\;j}\epsilon^{ij}\\
\label{eq:yukawa4}
\mathcal{L}_\mathrm{Y_{S_3}} \supset && -y^{1}_{S_3} Q_{L}^{i}L_{L}^{k}\epsilon^{ij} (\vec{\tau}.\vec{S}_3)^{jk} + y^{2}_{S_3}\epsilon^c Q_{L}^{i}Q_{L}^{k}\epsilon^{ij} (\vec{\tau}.\vec{S}_3)^{\dagger\;jk}
\end{eqnarray}
where $\epsilon^c$ corresponds to the $SU(3)_C$ tensor; $i,j,k$ are $SU(2)_L$ indices. From Eqs. (\ref{eq:yukawa1}), (\ref{eq:yukawa2}), (\ref{eq:yukawa3}) and (\ref{eq:yukawa4}) it follows that $B$ and $L$ can not uniquely be defined for $S_1$, $S'_1$, $\tilde{S}_1$ and $S_3$ in the presence of all allowed interactions; however, the quantum number $B-L$ can be uniquely defined for all the leptoquarks and $(B-L)_{S_1,S'_1,S_3}=+2/3$, $(B-L)_{\tilde{S}_1}=-2/3$, $(B-L)_{\tilde{R}_2,R_2}=+4/3$. From a symmetry point of view, in the presence of these scalar leptoquarks as new physics beyond the SM this simple observation hints towards $B-L$ as a fundamental local or global symmetry, and naturally motivates UV embedding of scalar leptoquarks in frameworks like $SO(10)$ or $G_{\text{Pati-Salam}}$, which can manifest $B-L$ as a local symmetry at a high scale. Using the $B-L$ assignments for the leptoquark states it can be readily verified from Eq. (\ref{eq:trilinear}) that each of the terms carry a $B-L$ charge $-2$ and consequently together with the Yukawa interaction given in Eqs. (\ref{eq:yukawa1}), (\ref{eq:yukawa2}), (\ref{eq:yukawa3}) and (\ref{eq:yukawa4}) they can give rise to $B-L$ violating proton decay operators given in Eq. (\ref{dim7}) with $\Delta(B-L) = +2$ \footnote{Note that one can also have $d=12$ nucleon decays in the presence of trilinear couplings involving three scalar leptoquarks \cite{Hambye:2017qix} and trilinear couplings involving a triplet Higgs scalar, a singlet and a triplet leptoquark \cite{Gu:2011pf}.}. 
\begin{figure}[t!]
\centering
	\includegraphics[scale=0.8]{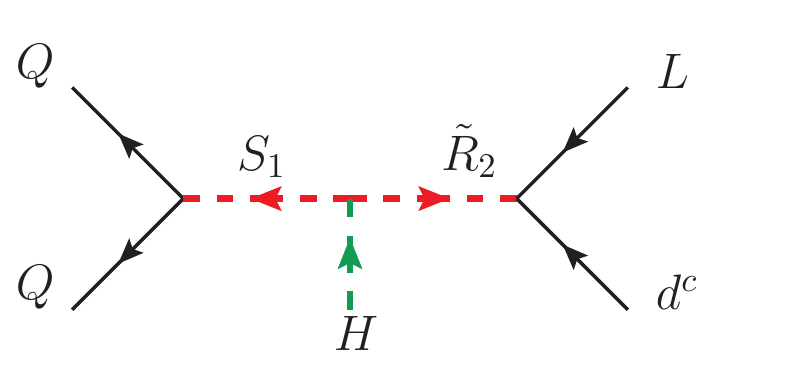}
	\includegraphics[scale=0.8]{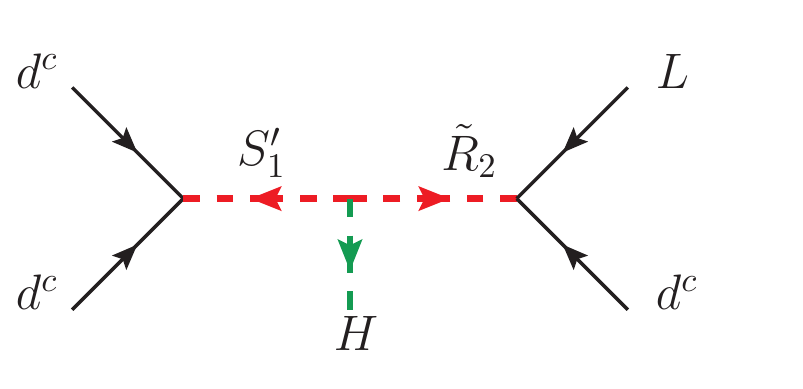}
	\includegraphics[scale=0.8]{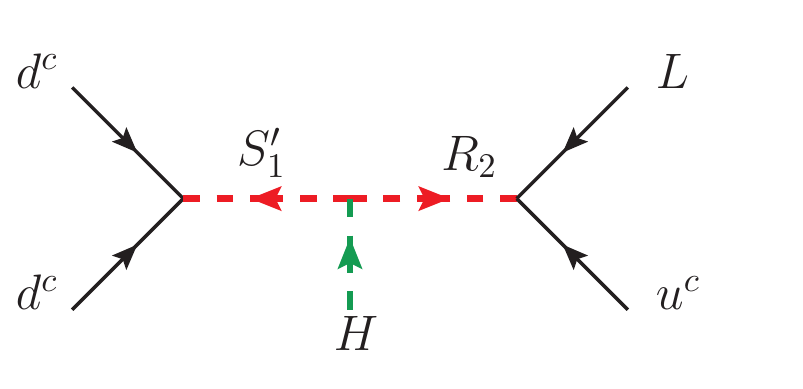}
    \includegraphics[scale=0.8]{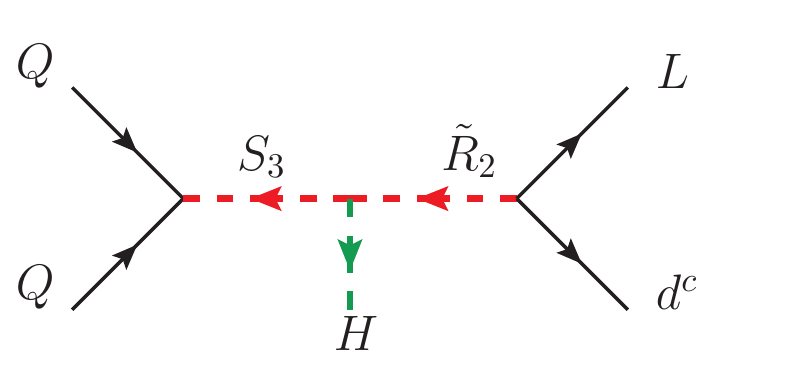}
	\includegraphics[scale=0.8]{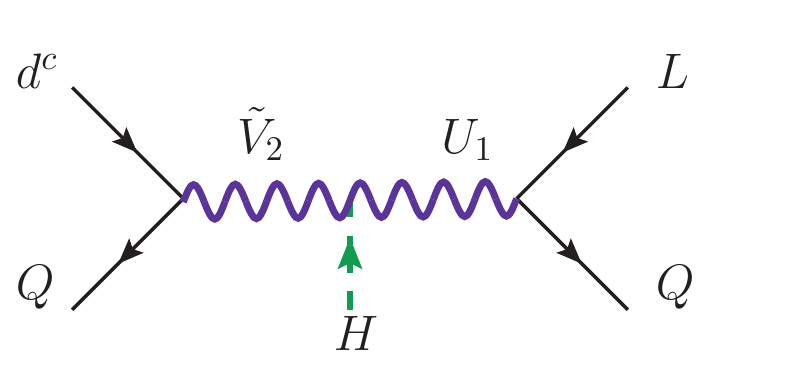}
	\includegraphics[scale=0.8]{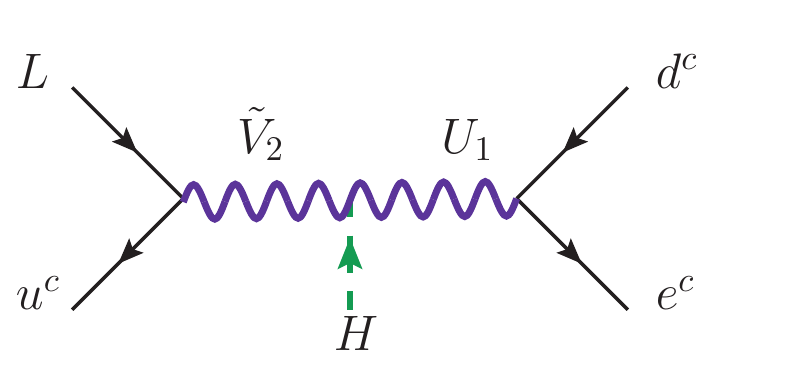}
	\caption{Some representative diagrams realising the effective $B-L$ violating $d=7$ operators induced by the trilinear leptoquark couplings.}
	\label{pdd}
\end{figure}
The vector leptoquarks $\tilde{V}_2:(\bar{3},2,-1/6)$ and  $U_1:(3,1,2/3)$ can also induce the $d=7$ operators of Eq. (\ref{dim7}) through the effective coupling
\begin{eqnarray}
\label{eq:vec}
\mathcal{L}_\mathrm{vector} \supset&& - \chi_1 \tilde{V}_2 U_1 H^{\dagger} +\textrm{h.c.}\; ,
\end{eqnarray}
where the vector leptoquarks couple to the SM fields through the interactions
\begin{eqnarray}
\label{eq:main_U_1}
\mathcal{L}_{U_1,\tilde{V}_2} &\supset &+x_{1} \bar{Q}_{L} \gamma^\mu U_{1,\mu} L_{L} + x_{2} \bar{d}^{c}_L \gamma^\mu U_{1,\mu}^{\dagger} e_{L}^{c}
+ x_{3}\bar{u}^{c}_L \gamma^\mu U_{1,\mu}^{\dagger} \nu_{L}^{c}+
 +\tilde{x}_{1}\bar{u}_{L}^{c\,i} \gamma^\mu \tilde{V}^{j}_{2,\mu} \epsilon^{ij}L_{L}^{i} \nonumber \\
 &&+\tilde{x}_{2}\bar{Q}_{L}^{i} \gamma^\mu \epsilon^{ij} \tilde{V}^{\dagger\,j}_{2,\mu}  \nu^{c}_{L}+\tilde{x}_{3}\bar{d}^{c}_{L} \gamma^\mu \tilde{V}^{\dagger}_{2,\mu} Q_{L} +\textrm{h.c.}\;.
\end{eqnarray}
The couplings in Eq. (\ref{eq:main_U_1}) leads to the $B-L$ assignments $(B-L) = (4/3, 2/3)$ for $(U_1,\tilde{V}_2)$. Consequently, Eq. (\ref{eq:vec}) together with the Yukawa interaction given in Eq. (\ref{eq:main_U_1})) lead to $\Delta(B-L)=2$ nucleon decay. In Fig. 1. we show some representative diagrams inducing $d=7$ $B-L$ violating nucleon decay operators in the presence of the scalar and vector leptoquark couplings discussed above. For illustration purposes we will consider the simple case of partial lifetime for the decay mode $n \rightarrow e^{-}\pi^{+}$. The lifetime for this decay mode induced by the trilinear coupling involving scalar leptoquarks $P_{1,2}$ and $H$, can be estimated by \cite{Nath:2006ut}
\begin{eqnarray}
\Gamma_s(n \rightarrow e^-\pi^+) \approx \frac{|Y^*_{P_1} Y_{P_2}|^2}{64 \pi}G^2 \frac{\beta_H^2 m_p}{f_\pi^2}\frac{\lambda^2 v^2}{M^4_{P_1}M^4_{P_2}},
\label{decayrate}
\end{eqnarray}
where $\lambda$ correspond to the dimension-ful $B-L$ violating trilinear coupling of $P_{1,2}$ and $H$; $Y_{P_1}$ and $Y_{P_2}$ are the Yukawa couplings for the scalar leptoquarks $P_1$ and $P_2$. $v\equiv \langle H^{0}\rangle=174$ GeV, $\beta_{H}\simeq 0.012\, {\text{GeV}}^{3}$ is the nucleon decay matrix element and $G\equiv(1+D+F)\simeq 1.3$ where the $D$ and $F$ correspond to contributions from chiral Lagrangian factors. The decay rate induced by the effective coupling involving vector leptoquarks $U_1,\tilde{V}_2$ and $H$ given in Eq. (\ref{vecdecayrate}) is given by
\begin{eqnarray}
\Gamma_v(n \rightarrow e^-\pi^+) \approx \frac{|x^*_{U_1} x_{\tilde{V}_2}|^2}{64 \pi} G^2 A \frac{|\bar{\alpha}_H|^2  m_p}{f_\pi^2}\frac{\chi_1^2 v^2}{M^4_{T_1}M^4_{T_2}}.
\label{vecdecayrate}
\end{eqnarray}
Here $x_{U_1}$ and $x_{\tilde{V}_2}$ are the couplings of $U_1$ and $\tilde{V}_2$ with SM fields, $A=|A_L|^{2}R$ with $R=[(A_{SR}+A_{SL})(1+|V_{ud}|^{2})^2]$ for $SO(10)$ where $A_L\sim 1.25$ and $A_L A_{SL}=A_L A_{SR}\sim 3.2$ \cite{Parida:2014dba,Bertolini:2013vta}. $\alpha_{H}=\bar{\alpha}_{H}G\simeq 0.012\, {\text{GeV}}^{3}$.

\section{UV completion and origin of $B-L$ violating couplings}{\label{sec3}}
At an effective level the computation of the decay rates given in Eqs. (\ref{decayrate}) and (\ref{vecdecayrate}) depends on the effective mass scales corresponding to the coupling coefficient of the $B-L$ violating effective interaction involving two leptoquarks and $H$, and the leptoquark mass scales. In this section we will show that one can have observable $B-L$ violating nucleon decay rates for some of the leptoquarks with mass as low as TeV scale, which is very interesting because the explanation the $B$-decay anomalies with scalar (vector) leptoquarks require the masses of such leptoquarks to be at around the TeV scale, and such mass scales of the leptoquarks will be probed by the direct searches at the LHC in the near future. Moreover, if the collider searches find leptoquark new physics at around TeV scale then the search for $B-L$ violating nucleon decays also provides an unique opportunity to indirectly probe $B-L$ breaking scale far beyond the reach of the current collider searches.

To discuss the realisations of the effective leptoquark couplings in the UV completion $SO(10)$, $G_{\text{Pati-Salam}}\equiv SU(2)_L \times SU(2)_R \times SU(4)_c$ and $SU(5)$, it is convenient to write down the decompositions of $SO(10)$ multiplets under the smaller subgroup representations. Under $G(2,2,4) \equiv SU(2)_L \times SU(2)_R \times SU(4)_C$ the $SO(10)$ multiplets decompose as
\begin{eqnarray}
10 &=& (2,2,1) + (1,1,6) \nonumber \\
16&=&(2,1,4)+(1,2,\overline{4}) \nonumber \\
120 &=& (2,2,1) + (1,1,10)+(1,1,\overline{10})+(3,1,6)+(1,3,6)+(2,2,15)\nonumber \\
126 &=& (1,1,6) + (3,1,10) + (1,3,\overline{10}) + (2,2,15)\nonumber \\ 
210 &=& (1,1,1) + (1,1,15) + (3,1,15) +  (1,3,15) + (2,2,6) + (2,2,10)+ (2,2,\overline{10}) ~.
\end{eqnarray}
Under the $G(5,1) \equiv SU(5) \times U(1)$ subgroup various multiplets decompose as follows:
\begin{eqnarray}
10 &=& 5(2) + \overline{5}(-2) \nonumber \\
16 &=& 1(-5) + \overline{5}(3) + 10(-1) \nonumber \\
120 &=& 5(2) + \overline{5}(-2) + 10(-6) + \overline{10}(6) + 45(2) + \overline{45}(-2)\nonumber \\
126 &=& 1(-10) + \overline{5}(-2) + 10(-6) + \overline{15}(6) + 45(2) + \overline{50}(-2) \nonumber \\ 
210 &=& (1,1,1) + (1,1,15) + (3,1,15) +  (1,3,15) + (2,2,6) + (2,2,10)+ (2,2,\overline{10}) ~.
\end{eqnarray}
 The effective $B-L$ violating couplings can naturally arise in GUT theories like $SO(10)$ and $G_{\text{Pati-Salam}}\equiv SU(2)_L \times SU(2)_R \times SU(4)_c$ where $B-L$ is a part of the local gauge symmetry. In such theories the effective $B-L$ violating couplings given in Eqs. (\ref{eq:trilinear}) and (\ref{eq:vec}) can arise when the local $B-L$ symmetry is broken by giving a vacuum expectation value to the SM singlet field $\Delta^{c}$ carrying $B-L=-2$. $\Delta^{c}$ is present in the $\overline{126}_H$ multiplet of $SO(10)$ and corresponds to $(1,3,\overline{10})$ under $G_{\text{Pati-Salam}}$ \cite{Babu:2012vb,Babu:2012iv}. Note that in SO(10) GUT $\Delta^{c}$ can also generate large Majorana masses for the right handed neutrinos through the couplings of the form $\nu^{c}\nu^{c}\Delta^{c}$. 
 
 In Table II we summarise the possible realisations of the relevant effective trilinear couplings given in Eq. (\ref{eq:trilinear}) in the context of $SO(10)$, $G_{\text{Pati-Salam}}$, and $SU(5)$ UV completion. The effective trilinear coupling $\tilde{R}_{2}^{\dagger} H S^{\dagger}_1$ can arise from the quartic coupling $(2,2,15) \cdot (2,2,15) \cdot (1,1,6) \cdot (1,3,\overline{10})$ under $G_{\text{Pati-Salam}}$ where $\tilde{R}_{2}^{\dagger}\subset (2,2,15)$, $S^{\dagger}_1 \subset (1,1,6)$, $H \subset (2,2,15)$ and $\Delta^{c}\subset (1,3,\overline{10})$. In $SO(10)$ such quartic coupling can be realised as $(126)^4$, $(126)^2(\overline{126})^2$ and $(126)^2(\overline{126}\, 10)$. The effective trilinear couplings $\tilde{R}^{\dagger}_{2} H^{\dagger} {S'}_1^{\dagger}$ and $R_{2}^{\dagger} H {S'}_1^{\dagger}$ can arise from the $G_{\text{Pati-Salam}}$ quartic term $(2,2,15) \cdot (2,2,15) \cdot (1,3,6) \cdot (1,3,10)$, where ${S'}_{1}^{\dagger}\subset (1,3,6)$ and $R_2\subset (2,2,15)$. In $SO(10)$ such quartic coupling can be realised by $(126)^2(120)^2$ and $(126\, \overline{126})(120)^2$. The quartic term $(126)^2(120)^2$ also contains the piece $(2,2,15) \cdot (2,2,15) \cdot (3,1,6) \cdot (1,3,10)$ under $G_{\text{Pati-Salam}}$ which can give the effective trilinear coupling $\tilde{R}_{2} S_{3} H^{\dagger}$ with $S_{3}\supset (3,1,6)$. 
 \begin{table}[h]
\centering
\begin{tabular}{|c|c|c|c|c|}
\hline
$G_{\text{SM}}$ effective coupling  &   $SO(10)$ & $G_{\text{Pati-Salam}}$ & $SU(5)$\\
\hline 
$\tilde{R}_{2}^{\dagger} H S^{\dagger}_1$   &  $(126)^4$; $(126)^2(\overline{126})^2$; & $(2,2,15) \cdot (2,2,15) \cdot $ & $\overline{15}\, 45\, 45 (24)$; $5\, \overline{10}\, 45 (24)$;\\
  &   $(126)^2(\overline{126}\, 10)$ & $(1,1,6) \cdot (1,3,\overline{10})$  & $\overline{5}\, \overline{5}\, 15(24)$; $\overline{5}\, \overline{5}\, 10\, 24$;\\
    &   &  & $5\, \overline{15}\, 45\, 24$\\
  \hline
$\tilde{R}^{\dagger}_{2} H^{\dagger} {S'}_1^{\dagger}$; $R_{2}^{\dagger} H {S'}_1^{\dagger}$   & $(126)^2(120)^2$;    & $(2,2,15) \cdot (2,2,15) \cdot $  & $5\, 10\, 10(24)$; $5\, 10\, 15\, 24$;\\
    & $(126\, \overline{126})(120)^2$  & $(1,3,6) \cdot (1,3,10)$ & $\overline{5}\, 10\, \overline{45} (24)$; $\overline{5}\, 10\, \overline{50}\, 24$\\
\hline
$\tilde{R}_{2} S_{3} H^{\dagger}$  &  $(126)^2(120)^2$ & $(2,2,15) \cdot (2,2,15) \cdot$  & $\overline{15}\, 45\, 45 (24)$; $5\, \overline{10}\, 45 (24)$;\\
 &   & $ (3,1,6) \cdot (1,3,10)$  & $5\, \overline{15}\, 45\, 24$\\
  \hline
\end{tabular}
\caption{\label{tab:2} Transformation properties of scalar leptoquarks under the SM gauge group $G_{\text{SM}}\equiv SU(3)_c \times SU(2)_L \times U(1)_Y $ and the relevant $SU(5)$, $G_{\text{Pati-Salam}}\equiv SU(2)_L \times SU(2)_R \times SU(4)_c$, $SO(10)$ representations.}
\end{table} 
 The effective Yukawa couplings given in Eqs. (\ref{eq:yukawa1}), (\ref{eq:yukawa2}), (\ref{eq:yukawa3}) and (\ref{eq:yukawa4}) can arise from the Yukawa couplings \cite{Aulakh:2004hm,Aulakh:2006hs} $16_i 16_j 10_H$, $16_i 16_j \overline{126}_H$ and $16_i 16_j 120_H$ in $SO(10)$ GUT theory, where the Yukawa couplings corresponding to the Higgs multiplets $10_H$ and $\overline{126}_H$ are symmetric and corresponding to the Higgs multiplet $120_H$ is antisymmetric. The terms relevant to the generation of the $d=7$ operators can be obtained decomposing the $SO(10)$ invariants
\begin{eqnarray}
{\cal L}(16_i 16_j 10_H) &=& h^{(10)}_{ij} \left[ (u^c_i Q_j + \nu^c_i L_j)\, h - (d^c_i Q_j + e^c_i L_j)\, h^{\dagger} +
\left( \frac{\epsilon}{2} Q_i Q_j + u^c_i e^c_j - d^c_i \nu^c_j \right)S_1^{\dagger} \right. \nonumber \\
&& \left. + \left(\epsilon u^c_i d^c_j + Q_i L_j  \right) S_1 \right],
\label{Yuk10}
\end{eqnarray}
\begin{eqnarray}
{\cal L}(16_i 16_j 120_H) &\supset& g^{(120)}_{ij}\left[(d_i Q^j+e^c_iL_j)\,h^{\dagger} - (u^c_i Q_j + \nu^c_i L_j)\, h - \sqrt{2} Q_i L_j\, S_1 \right. \nonumber \\
&& \left. - \sqrt{2} (u^c_i e^c_j - d^c_i \nu^c_j)\,S_1^{\dagger} -\frac{i}{\sqrt{3}}(d^c_i Q_j-3 e^c_iL_j)\, h^{(2)\dagger}  + \frac{i}{\sqrt{3}}
(u^c_i Q_j-3 \nu^c_i L_j)\, h^{(2)}  \right. \nonumber \\
&& \left. -2 e^c_i Q_j \,R_2^{\dagger}+ 2 \nu^c_i Q_j\, \tilde{R}_2^{\dagger}-2 d^c_iL_j \, \tilde{R}_2 + 2 u^c_i L_j \, R_2 \right. \nonumber \\
&& \left. - i\, \epsilon\, d^c_i d^c_j \, {S'}_1+ 2 \,i\, u^c_i \nu^c_j\, {S'}_1^{\dagger} + \sqrt{2}\, i\, \epsilon\, d^c_i u^c_j \, S_1^{(2)} + \sqrt{2}\, i\,
(d^c_i \nu^c_j-e^c_i u^c_j)\, S_1^{(2)\dagger} \right. \nonumber \\
&& \left. -\frac{\epsilon}{\sqrt{2}}Q_i Q_j S_3^{\dagger} - \sqrt{2}\, Q_i L_j S_3 -2 \,i\, d_i^c\,  e^c_j\, \tilde{S}_1^{\dagger} + i\, \epsilon\,
 u_i^c\, u_j^c\, \tilde{S}_1 \right].
\label{Yuk120}
\end{eqnarray}
\begin{eqnarray}
{\cal L}(16_i 16_j \overline{126}_H) &\supset& f^{(126)}_{ij} \left[(u^c_iQ_j - 3 \nu^c_i L_j)\,h - (d^c_iQ_j-3 e^c_iL_j)\,h^{\dagger} \right. \nonumber \\
&& \left.  + \sqrt{3}i\left(\frac{\epsilon}{2} Q_i Q_j - u^c_i e^c_j+ \nu^c_i d^c_j\right) S_1^{\dagger}
+ \sqrt{3} i (Q_i L_j - \epsilon u^c_i d^c_j)\,S_1 \right. \nonumber \\
&& \left.  + \sqrt{6}(d^c_i \nu^c_j + u^c_i e^c_j)\, S_1^{(2)\dagger}
+  2 \sqrt{3} i\, d^c_i\, L_j\, \tilde{R}_2  - 2 \sqrt{3} i\, \nu^c_i\, Q_j\, \tilde{R}_2 ^{\dagger} + 2 \sqrt{3} \, u^c_i \,\nu^c_j\,{S'}_1^{\dagger} \right. \nonumber \\
&& \left.- 2 \sqrt{3} i\, u_i^c \, L_j \, R_2+ 2 \sqrt{3} i\,  e_i^c \,Q_j \,R_2^{\dagger} -2 \sqrt{3}\, d_i^c\, e_j^c\, \tilde{S}_1^{\dagger} +
\sqrt{6} i \,Q_i\, L_j \,S_3 \right]~,
\label{Yuk126}
\end{eqnarray}
 where parenthesised superscripts have been used to distinguish more than one fields with the same quantum numbers appearing in some decompositions. In principle, after the GUT symmetry breaking various such subfields with the same SM quantum number would mix among themselves. The $h (1,2,1/2)$ and $h^{\dagger} (1,2,-1/2)$ fields from $10_H$ and $\overline{126}_H$ as well as any other Higgs multiplet with the same quantum number would mix to give a SM Higgs doublet which is a linear combination of all such fields. In $SU(5)$ GUT theory, the effective trilinear $B-L$ violating coupling e.g. $\tilde{R}_{2}^{\dagger} H S^{\dagger}_1$ can correspond to a number of possibilities as shown in Table \ref{tab:2}. Considering one example of such possibility $\overline{15}\, 45\, 45$ where $\tilde{R}_{2}^{\dagger}\subset \overline{15}$, $S^{\dagger}_1 \subset 45$, $H \subset 45$; one can embed such coupling as a quartic coupling $(126)^4$ in $SO(10)$, which corresponds to $\overline{15}(6) \cdot 45(2) \cdot 45(2) \cdot 1(-10)$ under the intermediate symmetry subgroup $SU(5)\times U(1)$. The vector leptoquarks $\tilde{V}_2$ and $U_1$ can also be embedded in $SO(10)$ framework and the covariant derivative of the $126_H$ multiplet would contain the quartic term $\tilde{V}_2 U_1 H^{\dagger} \Delta^c$ which leads to the effective term in Eq.(\ref{eq:vec}).

 \section{Connection to $B$-decay anomalies and observable $B-L$ violating nucleon decay rates}{\label{sec4}}
 A particularly interesting solution to both $R_{K^{(*)}}$ and $R_{D^{(*)}}$ anomalies is to have $R_2$ and $S_3$ leptoquarks at around TeV scale. The relevant effective Lagrangian for the charged current anomalies $R_{D^{(*)}}$ is given by
\begin{equation}
\begin{split}
{\cal L}^{d \to u \ell \bar \nu}_{\mathrm{eff}} = -\frac{4 \, G_F}{\sqrt{2}} &V_{ud}\big[  (1+g_{V_L}) 
(\bar{u}_L \gamma_\mu d_L)(\bar{\ell}_L \gamma^\mu \nu_{L})+ g_{S_L}(\mu)\, (\bar{u}_R  d_L)(\bar{\ell}_R \nu_{L}) \\
&+ g_T(\mu)\, (\bar{u}_R  \sigma_{\mu \nu} d_L) (\bar{\ell}_{R} \sigma^{\mu \nu}\nu_L)
\big]\,.
\label{eq:semilep}
\end{split}
\end{equation}
Interestingly, the relevant effective Wilson coefficients can get significant contributions from $R_2$ to resolve $R_{D^{(*)}}$ anomalies for $m_{R_2} \sim 1$ TeV after taking into account the constraints from the measurements of $R_{D^{(\ast)}}^{\mu/e} = \mathcal{B}(B\to D^{(\ast)}\mu \bar{\nu})/\mathcal{B}(B\to D^{(\ast)} e \bar{\nu})$,  $\mathcal{B}(B\to \tau \bar{\nu})$ and the kaon LFU ratio $R^K_{e/\mu}= \Gamma(K^-\to e^- \bar{\nu})/\Gamma(K^-\to \mu^- \bar{\nu})$. On the other hand, $S_3$ can contribute to $b\to s$ (semi-)leptonic transition to explain the anomalous data, but can give a negligibly small contribution to $R_{D^{(\ast)}}$ once the constraints from various favour violating processes are taken into account. The standard left-handed effective Hamiltonian for the $b\to s$ (semi-)leptonic transition is given by
\begin{equation}
\mathcal{H}^{b\to s l l}_{\mathrm{eff}} = -\dfrac{4 G_F \lambda_t }{\sqrt{2}}  \sum_{i=7,9,10} C_i(\mu)\mathcal{O}_i(\mu)\,,
\end{equation}
where $\lambda_t = V_{tb}V_{ts}^\ast$. The most relevant operators are
\begin{equation}
  \mathcal{O}_{9(10)} = \dfrac{e^2}{(4\pi)^2} \,\big{(}\bar{s}_L\gamma_\mu  b_L\big{)} \big{(}\bar{l}\gamma^\mu (\gamma^5)l \big{)}\,.
\end{equation}
$S_3$ can contributes at the tree-level inducing $\delta C_9^{\mu\mu} = - \delta C_{10}^{\mu\mu}$ ~\cite{Dorsner:2017ufx}. For a TeV scale $S_3$ one can address the $R_{K^{(\ast)}}$ anomalies while being consistent with various flavour violating constraints such as $\mathcal{B}(B_s\to \mu\mu)$, $B\to K^{(\ast)}\nu\nu$, $B_s-\bar{B}_s$ mixing amplitude etc. Using various available experimental measurements of $R_{K^{(*)}}$ in different high and low $q^2$ bins and the available experimental measurements on the angular
observables for $b\rightarrow s \mu \mu$ and $b\rightarrow s ee$, and
the latest average for BR$(B_s\rightarrow \mu\mu)$ one finds the global fit
ranges \cite{Hati:2019ufv}
\begin{eqnarray}
-0.50 \,(-0.58) &\geq  C_{9,sb}^{\mu\mu}=-C_{10,sb}^{\mu\mu} &\geq (-0.83)\, -0.91\,,\nonumber\\
0.00 \,(-0.12) &\geq  C_{9,sb}^{ee}=-C_{10,sb}^{ee} &\geq (-0.45)\, -0.55\,,
\end{eqnarray}
at the $2\,\sigma$ ($1\,\sigma$) level \footnote{Note that due to RG-running a large leptoquark coupling to third generation lepton  can potentially induce a non-negligible lepton-universal contribution to $b \to s \ell \ell$ transitions via a $\log$-enhanced anapole photon penguin contribution, as noted in Ref.~\cite{Crivellin:2018yvo}.}.

The possibility of accommodating such a scenario for $S_3$ leptoquark in the context of $SU(5)$ GUT completion has been discussed in Ref. \cite{Dorsner:2012nq,Becirevic:2018} and it has been explicitly pointed out that the term $10_i\,10_j\, 45_H$ gives rise to the di-quark coupling for $S_3$ and consequently, such a coupling must be absent to avoid constraints from $d=6$ proton decay if $S_3$ leptoquark mass is around TeV scale. Here we will be interested in the $SO(10)$ breaking to the SM gauge group through the intermediate subgroup $G_{\text{Pati-Salam}}\equiv SU(2)_L \times SU(2)_R \times SU(4)_c$. From the $SO(10)$ GUT completion perspective, from Eq. (\ref{Yuk120}) we note that the coupling $16_i 16_j 120_H$ would lead to $d=6$ rapid proton decay due to the presence of both di-quark and leptoquark couplings of $S_3$ leptoquark and therefore is inconsistent with a TeV scale $S_3$ explaining $R_{K^{(\ast)}}$ anomalies. More interestingly, we note that in the coupling $16_i 16_j \overline{126}_H$ the possibility of such a rapid proton decay due to $d=6$ operator induced by $S_3$ is absent and furthermore, we also notice that $R_2$ (which can possibly explain resolve $R_{D^{(*)}}$ anomalies for $m_{R_2} \sim 1$ TeV) can be embedded in such a coupling without inducing any rapid nucleon decay modes. Therefore, we will focus on this scenario where the pair $S_3-\tilde{R}_2$ belongs to the $\overline{126}_H$ multiplet and have masses at around TeV scale not inducing both $d=6$ and $d=7$ nucleon decays in the absence of the di-quark coupling of $S_3$. However, from Eq. (\ref{Yuk126}) it follows that the coupling $16_i 16_j \overline{126}_H$ induce $S_1$ mediated $d=6$ nucleon decay and consequently the experimental searches for $d=6$ nucleon decay constrain their mass scales to be relatively large $M_{S_1}>10^{11}$GeV for a benchmark Yukawa coupling $Y\sim 10^{-3}$. On the other hand, another submultiplet of $\overline{126}_H$, $\tilde{R}_2$ is not subject to such constrains and can have TeV scale mass. Therefore, if the interactions of the pair $S_3-R_2$ resolving $R_{K^{(\ast)}}$ and $R_{D^{(*)}}$ anomalies simultaneously, are traced back to the $SO(10)$ invariant coupling $16_i 16_j \overline{126}_H$, then one can have observable $B-L$ violating $d=7$ nucleon decays due to the $\overline{126}_H$ submultiplet pair $S_1-\tilde{R}_2$.

As a benchmark point taking $M_{\tilde{R}_2}\sim 1$ TeV, $M_{S_1}\sim 10^{16}$ GeV, $y_{S_1,\tilde{R}_2}\sim 10^{-3}$ we find from Eq. (\ref{decayrate}) that the lifetime $\tau \approx 3\times 10^{33}$ yrs for $\lambda_2=10^{11}$ GeV, which is in the observable range of current and future experiments. Such a choice for the mass spectrum can be motivated by an intermediate symmetry $G_{\text{Pati-Salam}}\equiv SU(2)_L \times SU(2)_R \times SU(4)_c$, which fixes the scale of $B-L$ symmetry breaking and consequently fixes the scale of $\lambda_2$. Interestingly, when the intermediate symmetry is $G_{\text{Pati-Salam}}\equiv SU(2)_L \times SU(2)_R \times SU(4)_c$ in the $SO(10)$ embedding of the scalar leptoquarks, one obtains gauge coupling unification at around $10^{16}$ GeV for the intermediate symmetry scale $M_I=M_{G_{\text{Pati-Salam}}}\approx 10^{11}$ GeV \cite{Babu:1992ia,Bajc:2001fe,Bajc:2002iw,Fukuyama:2002ch,Goh:2003sy,Babu:2005ia,Bertolini:2006pe,Joshipura:2011rr}. 
  
 For vector leptoquarks we note that a TeV scale $U_1$ can address the anomalies in $b\rightarrow s\ell\ell$ and $b\rightarrow c\ell \bar{\nu}$ data simultaneously. However, the strong bound from charged lepton flavour violating decay processes can only be satisfied if $U_1$ couples to the SM fields trough a mixing with vector-like counterparts of the SM fields and moreover the first generation mixing is further suppressed to account for the bounds from $K_L\rightarrow \mu e$ and $K \rightarrow \pi \mu e$ data (e.g. by some additional flavour symmetry) \cite{Bordone:2017bld,Calibbi:2017qbu,Blanke:2018sro,Hati:2019ufv}. To account for such a suppression we parametrise $x_{U_1,\tilde{V}_2}=\beta g_U$ where $\beta \sim 10^{-3}$. Taking $M_{U_1}\sim 1$ TeV and $M_{\tilde{V}_2}\sim 10^{16}$ GeV (as it can also mediate $d=6$ nucleon decay) we obtain from Eq. (\ref{vecdecayrate}) the lifetime $\tau \approx 1.8\times 10^{34}$ yrs for the effective $B-L$ violating scale $\chi_1=10^{11}$ GeV, which is again in the observable range of current and future experiments searching for nucleon decays.

\section{Implications for baryogenesis and comments on neutrino masses}
\begin{figure}[h]
\centering
	\includegraphics[scale=1.0]{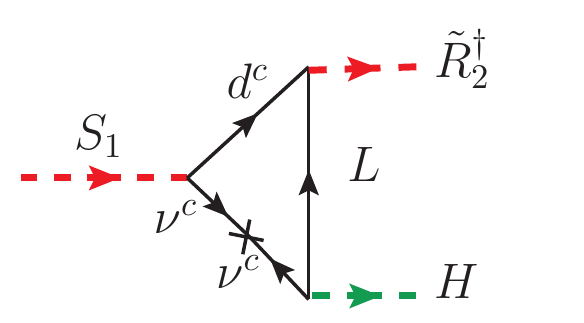}
	\includegraphics[scale=1.0]{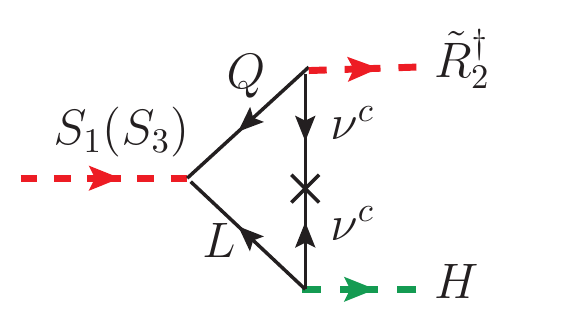}
	\includegraphics[scale=0.7]{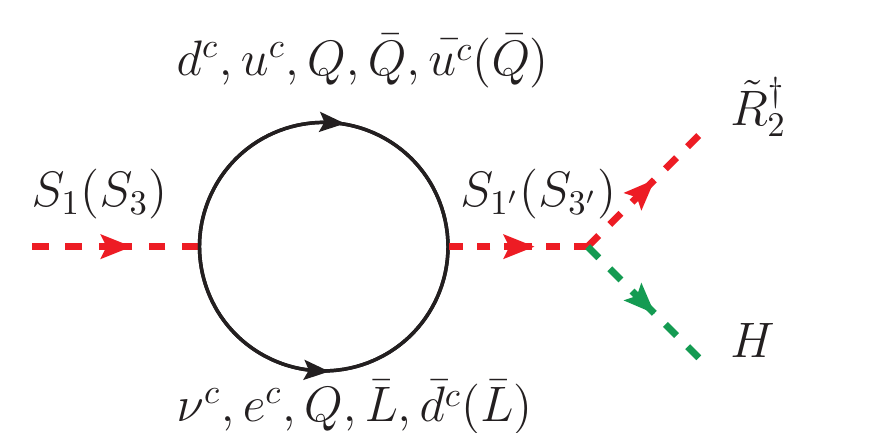}
	\includegraphics[scale=1.0]{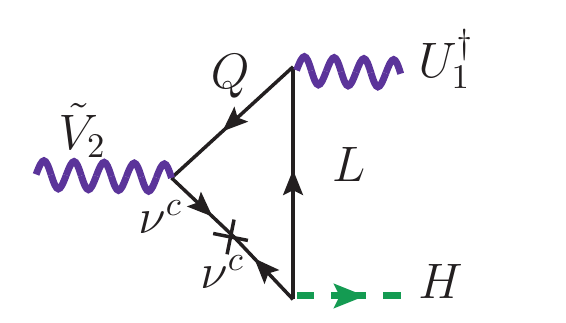}
	\caption{One loop corrections responsible for generating $(B-L)$ asymmetry in  $S_{1,3}$ and $\tilde{V}_2$ decays.}
	\label{baryo1}
\end{figure}

{\label{sec5}}
The $(B-L)$ violating effective couplings in Eqs. (\ref{eq:trilinear}) and (\ref{eq:vec}) open up a new possibility of realising high scale  or resonant baryogenesis \footnote{A complete list of trilinear couplings which can contribute to baryogenesis can be found  for example in Ref. \cite{Ma:1998hn}.}. Some of the decay modes which can give rise to high scale baryogenesis are $S_1\rightarrow \tilde{R}_2^{\dagger} H$, $S'_1\rightarrow \tilde{R}_2^{\dagger} H^{\dagger}$, $S'_1\rightarrow R_2^{\dagger} H$, with $S_1$ and $S'_1$ having GUT scale masses $\sim10^{16}$ GeV and $M_{\tilde{R}_2,R_2}\sim$ TeV. On the other hand, if $M_{S_3}>M_{\tilde{R}_2}$, with both of them having masses at around TeV scale then one can have resonant baryogenesis through the decay mode $S_3\rightarrow \tilde{R}_2^{\dagger} H$ in the presence of at least two generations of $S_3$ with nearly degenerate masses. For vector leptoquarks one can have high scale baryogenesis through the decay mode $\tilde{V_2}\rightarrow U_1^{\dagger} H$. In Fig. \ref{baryo1} we show the one loop diagrams which can interfere with the tree level decay modes of $S_{1,3}$ and $\tilde{V}_2$ to provide the CP violation. The computation of the final asymmetry is UV model dependent and must also take into account the potential gauge and Yukawa washout processes \cite{Gu:2007bw}. 

We will consider here the simplest case of the $(B-L)$--violating decay of the scalar $S_1$, which we shall assume to be very heavy with mass $\sim 10^{16}$~GeV. In a UV complete theory such as $SO(10)$ the leptoquark $S_1$ is in general a linear combination of fields from the decomposition of multiplets $10$ and $\overline{126}$ and $120$ fields. Firstly, we note that the decay $S_1\rightarrow \tilde{R}_2^{\dagger} H$ would violate $(B-L)$ by $2$ units, and because $H$ carries zero $(B-L)$ charge the final state has $(B-L) = -4/3$. On the other hand $S_1$ can also decay via the $B-L$ conserving fermionic modes $S_1 \rightarrow ff'$ with  $(B-L) = 2/3$ in the final state  \footnote{The interplay between the $(B-L)$--conserving decays and the $(B-L)$--preserving decays of leptoquarks are crucial for inducing generating the asymmetry as first pointed out in Refs. \cite{Babu:2012vb,Babu:2012iv}. }.

In the absence of more than one generation of $S_1$ only the top two diagrams in Fig. \ref{baryo1} contributes and the contribution coming from these diagrams are proportional to the Majorana masses for  $\nu^c$. Assuming the Majorana mass matrix of the $\nu^c$ fields to be diagonal and real, the asymmetry contribution coming from these two diagrams is given by
\begin{eqnarray}
\epsilon_{B-L}^{\text{v}} &=& \epsilon_{B-L}^{\text{I}}+\epsilon_{B-L}^{\text{II}}\nonumber\\
 &=& -\frac{f_r}{\pi} {\rm Im} \left[\frac{{\rm Tr}\{y_{S_1}^{1} \, y_{\tilde{R}_2}^{2\dagger}\,M_{\nu^c} \, F_{S_1,\nu^c, \tilde{R}_2}\, y_{\nu^c L H} )\} \, \lambda_1}{|\lambda_1|^2} \right] \nonumber\\
 &&+ \frac{f_r}{\pi} {\rm Im} \left[\frac{{\rm Tr}\{y_{S_1}^{5} \, y_{\tilde{R}_2}^{1\dagger}\,\, y_{\nu^c L H} \, M_{\nu^c}\, F_{ \tilde{R}_2,\nu^c,S_1}\}\, \lambda_1}{|\lambda_1|^2} \right]\nonumber\\
\label{asymb}
\end{eqnarray}
Here we define the Yukawa coupling matrix corresponding to the coupling $\nu^c L H$, as  $y_{\nu^c L H}$ and 
$M_{\nu^c}$ is the real diagonalised mass matrix of $\nu^c$ fields. ${\rm f_{r}}$ corresponds to the branching ratio
${\rm Br}(S_1^{\dagger} \rightarrow \tilde{R}_2 H^{\dagger})$.  
The function $F_{i,j,k}\equiv F(M_i,M_j,M_k)$ is defined as
\begin{equation}
F(M_i,M_j,M_k) = {\rm ln}\left(1+ \frac{M_i^2}{M_j^2}\right) + \Theta\left(1- \frac{M_j^2}{M_k^2}\right)\left(1- \frac{M_j^2}{M_k^2}\right),
\end{equation}
where  $\Theta$ stands for the step function, accounting for additional ways of cutting the
diagram for the case $M_{\nu^c} < M_{\tilde{R}_2}$.

In addition to the above contributions, another additional contribution can arise in any realistic UV complete model where there are more than one $S_1$ fields.  Denoting the heavier $S_1$ as $S_{1'}$ the contribution coming from the bottom-left diagram of Fig. \ref{baryo1} is given by \footnote{We assume $M_{S_1} - M_{S_{1^{\prime}}} \gg \Gamma_{S_1}$, so that there is no resonant enhancement for the decay.}
\begin{equation}
\epsilon_{B-L}^{s} = \frac{f_r}{\pi} {\rm Im} \left[\frac{{\rm Tr}\{y^{5\dagger}_{S_{1'}} \, y^{5}_{S_1}
\, G (S_1,S_{1^\prime},{\nu^c}) \}  (\lambda_{1^{\prime}})^*(\lambda_1)}{|\lambda_1|^2} \right],
\label{asymd}
\end{equation}
where the loop function $G(S_1,S_{1^\prime},{\nu^c})\equiv G(M_{S_1},M_{S_{1^{\prime}}},{\nu^c})$ is defined as
\begin{equation}
G(M_i,M_j,M_k) =  \left(\frac{M_i^2 - M_k^2}{M_i^2-M_{j}^2}\right) \Theta\left(1- \frac{M_k^2}{M_i^2}\right)\,\left(1- \frac{M_j^2}{M_i^2}\right)~.
\end{equation}
For the case where $S_1$ and $S_{1^{\prime}}$ are degenerate in mass, a resonant enhancement is possible.  For a prescription of such a case see for example Ref. \cite{Pilaftsis:2003gt}.

The branching ratio $f_r\equiv \Gamma(S_1^{\dagger} \rightarrow \tilde{R}_2 H^*) /(\Gamma(S_1^{\dagger} \rightarrow \tilde{R}_2 H^*) +\Gamma (S_1^{\dagger} \rightarrow ff))$ in the asymmetry  can be estimated using the partial decay widths
\begin{eqnarray}
\Gamma(S_1^{\dagger} \rightarrow \tilde{R}_2 H^*)  &=& \frac{|\lambda_1|^2}{8 \pi M_{S_1}}\left(1-\frac{M_{\tilde{R}_2}^2}{M_{S_1}^2}  \right),\nonumber\\
\Gamma (S_1^{\dagger} \rightarrow ff) &=& \frac{{\rm Tr}(y_{S_1}^{f\dagger} y_{S_1}^f) }{4 \pi} M_{S_1}.
\label{partial}
\end{eqnarray}

\begin{figure}[b!]
\centering
	\includegraphics[scale=0.50]{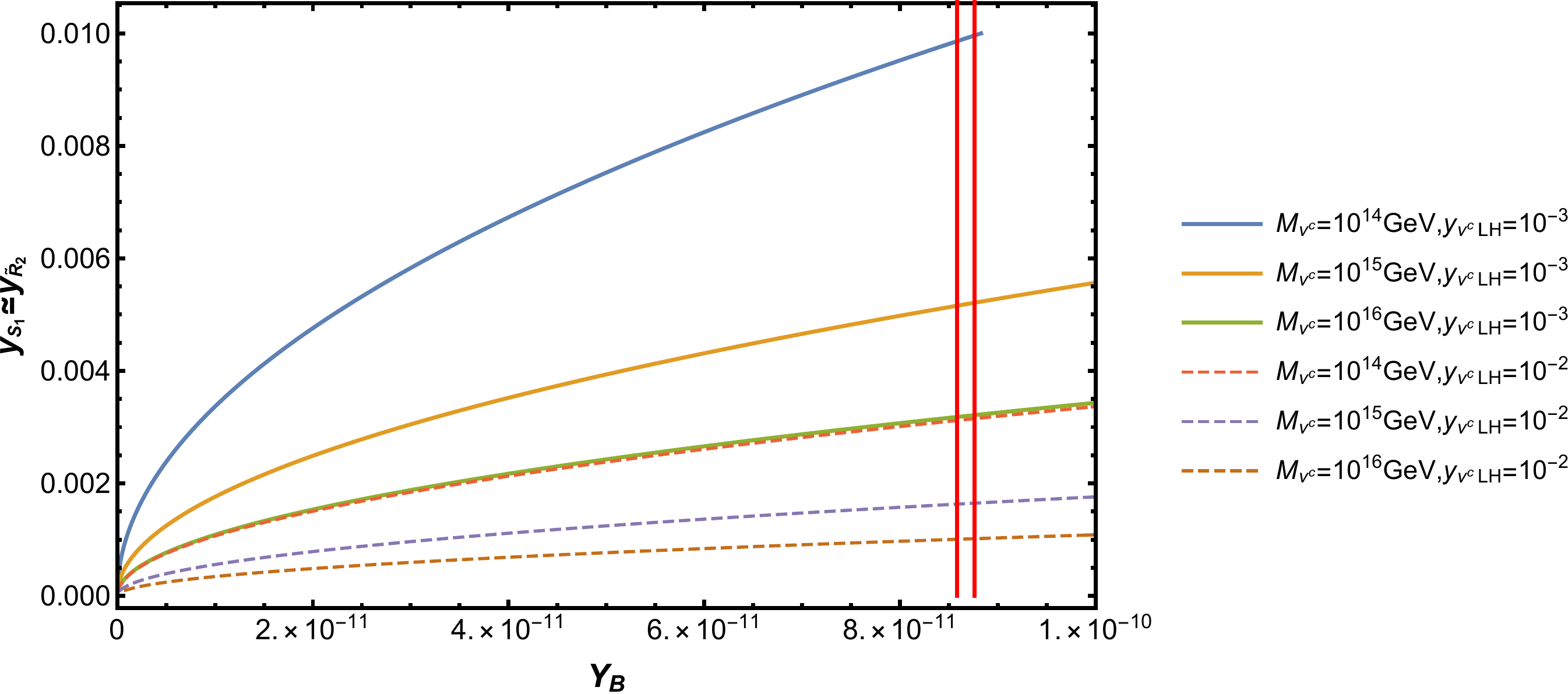}
    \caption{The dependence of the final baryon asymmetry on the leptoquark Yukawa couplings $y_{S_1}\sim y_{\tilde{R}_2}$, for different values of $M_{\nu^c}$ and $y_{\nu^c L H}$. the vertical red lines correspond to the current best fit for the baryon asymmetry to entropy ratio provided by the PLANK collaboration. We have chosen the benchmark values $M_{\tilde{R}_2}= 1$ TeV, $M_{S_1}= 10^{16}$ GeV, $\lambda=10^{11}$ GeV and $y_{S_1}^f = 10^{-3}$ for this plot.}
	\label{baryo2}
\end{figure}

\begin{figure}[htb!]
\centering
	\includegraphics[scale=0.50]{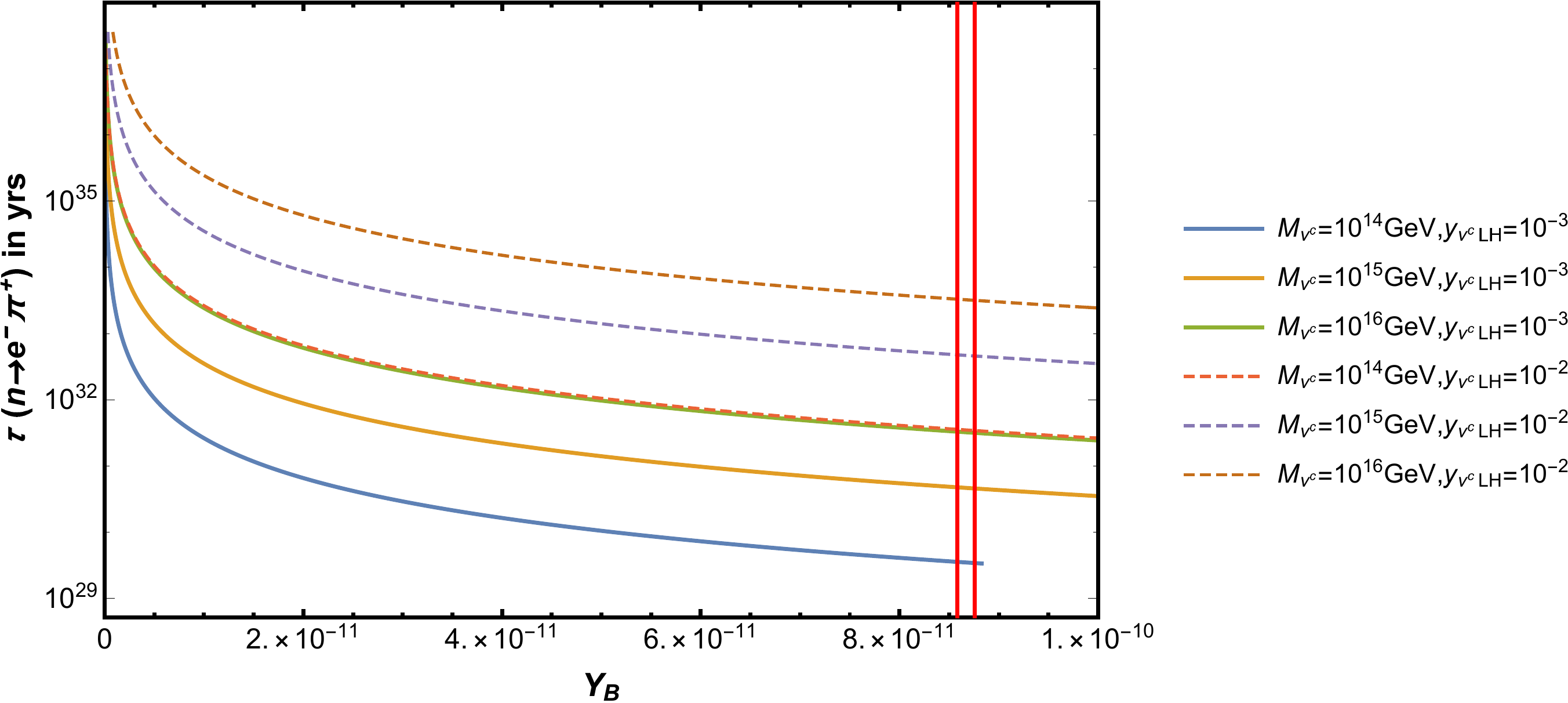}
    \caption{Correlation between the final baryon asymmetry and the partial lifetime for the decay mode $n \rightarrow e^{-}\pi^{+}$ for different values of $M_{\nu^c}$ and $y_{\nu^c L H}$ with the benchmark values $M_{\tilde{R}_2}= 1$ TeV, $M_{S_1}= 10^{16}$ GeV, $\lambda=10^{11}$ GeV and $y_{S_1}^f = 10^{-3}$.}
	\label{baryo3}
\end{figure}

The baryon asymmetry to entropy ratio $Y_B$ is given by
\begin{equation}
Y_B \equiv \frac{n_B - n_{\overline{B}}}{s} = \frac{\epsilon_{B-L}}{g_*} d~,
\label{eta}
\end{equation}
where $g_*$ is the total number of relativistic degrees of freedom and $d$ in Eq. (\ref{eta}) is the dilution factor taking into account the partial wash out processes, which can be obtained exactly by solving the Boltzmann equation. The dilution factor can be approximately estimated by
\begin{equation}
d\simeq \left\{\begin{array}{ll}
 1 & ~~~(K < 1) \\
 \frac{0.3}{K ({\rm ln}\,K)^{0.6}} & ~~~(K \gg 1),
\end{array} \right.
\end{equation}
where
\begin{equation}
K = \left. \frac{\Gamma(S_1^{\dagger} \rightarrow \tilde{R}_2 H^*)} {2{\rm H}}\right|_{T=M_{S_1}}.
\end{equation}
with ${\rm H}$ denoting the Hubble expansion rate given by
\begin{equation}
{\rm H} = 1.66\, g_*^{1/2} \frac{T^2}{M_{\rm Pl}}.
\end{equation}
In Fig. \ref{baryo2}, we show the dependence of the final baryon asymmetry on the couplings for $y_{S_1}\sim y_{\tilde{R}_2}$, for different values of $M_{\nu^c}$ and $y_{\nu^c L H}$, while choosing the benchmark values for rest of the parameters: $M_{\tilde{R}_2}= 1$ TeV, $M_{S_1}= 10^{16}$ GeV, $\lambda=10^{11}$ GeV and $y_{S_1}^f = 10^{-3}$. A direct correlation between the final baryon asymmetry and the lifetime for nucleon decay mode can also be obtained under the simplifying assumption that the leptoquark dominantly couples to first generation SM couplings \footnote{This assumption can always be relaxed to introduce non-trivial couplings of the leptoquark with second and third generation fermion couplings, therefore introducing a number of new paramers generalising $y_{S_1}$ and $y_{\tilde{R}_2}$ to matrices. }. In Fig. \ref{baryo3}, we show the correlation between the final baryon asymmetry and the partial lifetime for the decay mode $n \rightarrow e^{-}\pi^{+}$ for different values of $M_{\nu^c}$ and $y_{\nu^c L H}$ with the same benchmark values for the rest of the parameters as in Fig. \ref{baryo2}. It is important to note that various partial wash out processes can occur by means of various $(B-L)$ violating interactions of the right-handed neutrinos, after the $\nu^c$ fields acquire Majorana masses after $(B-L)$ breaking. Depending on the region of the parameter space, it may be desirable to have partial wash-out. On the other hand, any wash out effects can be inhibited by decoupling the $\nu^c$ fields at the same temperature as the $S_1$ field
($M_{\nu^c} \sim M_{S_1}$). 

Furthermore, the generation of baryon asymmetry can also be directly linked to the neutrino masses in the presence of $\Delta^{c}$ which can induce the effective $B-L$ violating couplings of the leptoquarks and generate large Majorana masses for the right handed neutrinos through a term $\nu^{c}\nu^{c}\Delta^{c}$. Alternatively, neutrino masses can also be generated through one loop corrections involving the $B-L$ violating vertices and up (down) type quarks in the loop \cite{Dorsner:2017wwn}. Thus depending on the mass scales and UV completed model these two contributions will compete to generate the correct neutrino masses, nevertheless, both these possibilities have the $B-L$ violating scale embedded in them which can be probed by the $B-L$ violating nucleon decay modes.


\section{Conclusion}{\label{sec6}}
In conclusion, we have shown that the effective $B-L$ violating couplings of scalar and vector leptoquarks can naturally induce dimension seven $B-L$ violating nucleon decay modes. The decay lifetime of such $B-L$ violating nucleon decay modes are sensitive to the ratio of effective coupling proportional to the $B-L$ breaking scale and the mass scales of the leptoquarks. Consequently, a discovery of a $d=7$ $B-L$ violating nucleon decay mode will certainly provide a strong evidence in the favor of leptoquark new physics and complimented by the upcoming results from the B-factories and the collider searches such observation can potentially probe the $B-L$ breaking scale as high as $10^{11}$ GeV, paving way for a new baryogenesis mechanism and giving interesting hints towards our understanding of neutrino masses. We have discussed some examples of possible UV completion for the relevant leptoquark interactions (in the context of GUT theories such as $SU(5)$, $G_{\text{Pati-Salam}}\equiv SU(2)_L \times SU(2)_R \times SU(4)_c$ and $SO(10)$) and have explored the possibility of having observable $B-L$ violating nucleon decay rates for some of the leptoquarks with mass as low as TeV scale. This is particularly very interesting because the explanation of the $B$-decay anomalies with scalar (vector) leptoquarks require the masses of such leptoquarks to be at around the TeV scale. Furthermore, if the upcoming collider searches find leptoquark new physics at around TeV scale then the search for $B-L$ violating nucleon decays will provide a unique opportunity to indirectly probe $B-L$ breaking scale far beyond the reach of the current collider searches and can have further implications for baryogenesis as we have discussed in the case of the simple example of scalar leptoquarks. We have also explored the correlation between the final baryon asymmetry and the lifetime for nucleon decay mode to this end. 
\section{Acknowledgements}
The work of C.H. is partly supported by the framework of the European Union's Horizon 2020 research and innovation programme under the Marie Sklodowska-Curie grant agreements No 690575 and No 674896. The work of U.S. is supported by
the J. C. Bose Fellowship, DST, India.

\end{document}